\documentclass{iaus}
\usepackage{graphics}
\makeindex

\newcommand{\MgII}{\hbox{{\rm Mg}\kern 0.1em{\sc ii}}}

\pubyear{2007}
\volume{241}
\pagerange{1--2}
\date{}
\jname{Stellar populations as bulding blocks of galaxies}
\editors{A. Vazdekis \& R. F. Peletier, eds.}

\begin{document}

\halftitle{INTERNATIONAL ASTRONOMICAL UNION\\
UNION ASTRONOMIQUE INTERNATIONALE}

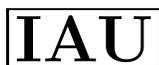
\title{STELLAR POPULATIONS \\
as BUILDING BLOCKS\\
of GALAXIES}

\where{PROCEEDINGS OF THE 241th SYMPOSIUM OF THE\\
INTERNATIONAL ASTRONOMICAL UNION\\
HELD IN LA PALMA, TENERIFE, SPAIN \\
DECEMBER 10-16, 2006}

\author{ALEXANDRE VAZDEKIS
\affil{Instituto de Astrof\'\i sica de Canarias, Tenerife, Spain} 
\and
        REYNIER PELETIER 
\affil{Kapteyn Astronomical Institute, Groningen, The Netherlands}
      }

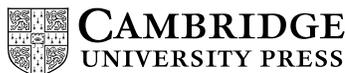
\makebooktitle

\begin{center}\vspace*{.25\baselineskip}
{\bf IAU SYMPOSIUM PROCEEDINGS SERIES}

{2006 EDITORIAL BOARD}

\bigskip

\small

{\normalsize\it Chairman}

K.A. VAN DER HUCHT,
IAU Assistant General Secretary                    \\
{\it SRON Netherlands Institute for Space Research \\
Sorbonnelaan 2,
NL-3584 CA Utrecht, The Netherlands}               \\
K.A.van.der.Hucht@sron.nl

\bigskip

{\normalsize\it Advisors}

O. ENGVOLD, IAU General Secretary,
{\it Institute of Theoretical Astrophysics, University of Oslo, Norway} \\
E.J. DE GEUS, {\it Netherlands Foundation for Research in Astronomy, Dwingeloo, The Netherlands} \\
M.C. STOREY, {\it Australia Telescope National Facility, Australia} \\
P.A. WHITELOCK, {\it South African Astronomical Observatory, South Africa}

\bigskip

{\normalsize\it Members}

IAUS233                                                                      \\
V. BOTHMER, {\it Universit\"ats Sternwarte, Georg-August-Universit\"at,
                      G\"ottingen, B.R. Deutschland}                         \\
IAUS234                                                                      \\
M.J. BARLOW, {\it Department of Physics and Astronomy, University College
                      London, London, UK}                                    \\
IAUS235                                                                      \\
F. COMBES, {\it LERMA, Observatoire de Paris, Paris, France}                 \\
IAUS236                                                                      \\
A. MILANI, {\it Dipartimento di Matematica, Universit\`a di Pisa, Pisa,
                       Italia}                                               \\
IAUS237                                                                      \\
B.G. ELMEGREEN, {\it IBM Research Division, T.J. Watson Research Center,
                       Yorktown Heights, NY, USA}                            \\
IAUS238                                                                      \\
V. KARAS, {\it Astronomical Institute, Academy of Sciences of the Czech
                      Republic, Praha, Czech Republic}                       \\
IAUS239                                                                      \\
F. KUPKA, {\it Max-Planck-Institut f\"ur Astrophysik, Garching-bei M\"unchen,
                      B.R. Deutschland}                                      \\
IAUS240                                                                      \\
W.I. HARTKOPF, {\it U.S. Naval Observatory, Washington D.C., USA}            \\
IAUS241                                                                      \\
A. VAZDEKIS, {\it Instituto de Astrof\'{\i}sica de Canarias, La Laguna,
                      Tenerife, Canary Islands, Spain}                       \\
\end{center}
\vfill\eject

\begin{titlepage}
\begin{flushleft}
\footnotesize
\textsc{c\ls a\ls m\ls b\ls r\ls i\ls d\ls g\ls e\ns 
u\ls n\ls i\ls v\ls e\ls r\ls s\ls i\ls t\ls y\ns p\ls r\ls e\ls s\ls s}\\

The Edinburgh Building, Cambridge CB2 2RU, UnitedKingdom\\
40 West 20th Street, New York, NY 10011--4211, USA\\
10 Stamford Road, Oakleigh, Melbourne 3166, Australia

\hbox{}

\noindent\copyright\ International Astronomical Union 2007\\

\hbox{}

This book is in copyright. Subject to statutory exception\\
and to the provisions of relevant collective licensing agreements,\\
no reproduction of any part may take place without\\
the written permission of the International Astronomical Union.

\hbox{}

\noindent First published 2007\\

\hbox{}

\noindent Printed in the United Kingdom at the University Press, Cambridge

\hbox{}

\noindent Typeset in System \LaTeXe

\hbox{}

\noindent {\it A catalogue record for this book is available from the British Library}

\hbox{}

\noindent{\it Library of Congress Cataloguing in Publication data}\vskip9\baselineskip

\hbox{}

\noindent{ISBN 0 521 57156 1 hardback}\\

\noindent{ISSN }
\end{flushleft}
\end{titlepage}

\title[Contents] {Contents}


\makeatletter
\def\@pnumwidth{2.4em}
\def\tocline#1#2#3{\medskip\vbox{
   \@dottedtocline{1}{\z@}{.4in}{#1}{#3}
   \smallskip
   \@authortocline{1}{.4in}{#2}}
}
\def\@authortocline#1#2#3{\ifnum #1>\c@tocdepth \else
  \vskip \z@ plus .2pt
  {\leftskip #2\relax \rightskip \@tocrmarg
   \parindent\z@\relax\@afterindenttrue
   \interlinepenalty\@M
   \leavevmode
   \@tempdimb\textwidth \advance\@tempdimb by-#2
   \advance\@tempdimb by-\@tocrmarg
   \parbox{\@tempdimb}{\pretolerance=10000\raggedright\small\sl#3}\par}\fi}
\newcounter{tocpart}
\newcounter{tocsect}[tocpart]
\def\sectline#1{\stepcounter{tocsect}\vskip 3.5ex
    {{\bf Section \Alph{tocsect}. #1}}\par\smallskip}
\def\partline#1{\stepcounter{tocpart}\vskip 5.5ex
    {\Large\bf Session \arabic{tocpart}. #1}\par}
\makeatother

\parindent 0pt

\pagenumbering{roman}
\setcounter{page}{5}
\markboth{Contents}{Contents}

.
\vskip 25truemm
\section*{\Large{Contents}}
\vskip 15truemm

\tocline{Preface}{}{xvii}
\tocline{Organising Committee}{}{xix}
\smallskip

\thispagestyle{plain}

\partline{Model Ingredients: Stellar Evolution Models}{}
\tocline{Stellar Evolutionary Models: challenges from observations of stellar systems}{Santi Cassisi}
{3}
\tocline{Populations of massive stars in galaxies, implications
for the stellar evolution theory}{Georges Meynet, Patrick Eggenberger and Andr\'e Maeder}
{13}
\tocline{On Stellar Models with Blanketed Atmospheres as Boundary Conditions}{Don A.~VandenBerg, Bengt Edvardsson, Kjell Eriksson,
 Bengt Gustafsson and Jason W.~Ferguson}
{23}
\tocline{Stellar Evolution Challenge}{A. Weiss, S. Cassisi, A. Dotter, 
Z. Han and Y. Lebreton}
{28}
\tocline{Thresholds for the Dust Driven Mass Loss from C-rich AGB Stars}{Lars Mattsson, Rurik Wahlin and Susanne H\"ofner}
{37}
\tocline{BaSTI - a library of stellar evolution models: updates and
applications}{A. Pietrinferni, S. Cassisi, M. Salaris, D. Cordier
 and F. Castelli}
{39}
\tocline{Are interpolations in metallicity reliable?}{L.Angeretti, G. Fiorentino  and L. Greggio}
{41}
\tocline{Low-mass stellar models with new opacity tables and varying $\alpha$-element enhancement factors}{A. Weiss, J. Ferguson, and M. Salaris}
{43}

\partline{Model Ingredients: Stellar Spectral Libraries}{}
\tocline{Libraries of synthetic stellar spectra -- or are we building palaces upon sand?}{Bengt Gustafsson, Ulrike Heiter and Bengt Edvardsson}
{47}
\tocline{A High Resolution $\alpha$-enhanced stellar Library for Evolutionary Population Synthesis}{Lucimara Martins and 
 Paula Coelho}
{58}
\tocline{Metallicity and age of M31 globulars from automated fits to theoretical spectra}{Ruth C. Peterson}
{63}
\tocline{Analysis of stellar populations with large empirical libraries at high 
spectral resolution}{Philippe Prugniel, Mina Koleva, Pierre Ocvirk, Damien Le Borgne and Caroline Soubiran}
{68}
\tocline{Abundances in the Galactic Bulge: evidence for fast chemical enrichment}
{M. Zoccali,
A. Lecureur, B. Barbuy, V. Hill, A. Renzini,
D. Minniti, Y. Momany, A. G\'omez 
and
S. Ortolani}
{73}
\tocline{Chemical Abundances in Metal-Rich Bulge-like Stars}{L. Pomp\'eia, B. Barbuy, M. Grenon and B. Gustafsson}
{78}
\tocline{Flux Calibration Issues}{Andrew J. Pickles}
{82}
\tocline{Spectra of bulge stars with known abundance ratios for population synthesis}{B. Barbuy, A. Alves-Brito, M. Zoccali, D. Minniti, V. Hill, 
A. Lecureur, A. G\'omez, S. Ortolani, A. Renzini, P. Coelho and L. Sodr\'e}
{87}
\tocline{Physical requirements for modeling stellar atmospheres according to the different spectral features 
observed}{L. Crivellari, O. Cardona
and E. Simonneau}
{91}
\tocline{Cross-checking reliability of some available stellar spectral libraries
using artificial neural networks }{Ranjan Gupta, S. Jotin Singh and Harinder P. Singh}
{93}
\tocline{Hubble's Next Generation Spectral Library}{Sara R. Heap and Don Lindler}
{95}
\tocline{A new stellar library in the K band for the empirical calibration of the CO
index}{E. M\'armol-Queralt\'o, N. Cardiel, A.~J. Cenarro, A.
Vazdekis, F. J. Gorgas, and R.~F. Peletier}
{97}
\tocline{New Empirical Fitting Functions of the Lick/IDS indices using MILES}{J.~M. Mart\'{\i}n-Hern\'andez, E. M\'armol-Queralt\'o,
J. Gorgas,
N. Cardiel, P. S\'anchez-Bl\'azquez,
A.~J. Cenarro, R.~F. Peletier, 
A. Vazdekis and J. Falc\'on-Barroso}
{99}
\tocline{Filling Gaps in Indo-US Stellar Spectral Library using Principal Component Analysis }{Harinder P. Singh, S. Jotin Singh, Ranjan Gupta and M. Yuasa}
{101}
\tocline{Spectral synthesis in the near UV (3000-4500 \AA)}{Rodolfo Smiljanic and Beatriz Barbuy}
{103}
\tocline{Towards a low metallicity carbon star spectral library}{R. Wahlin,
 L. Mattsson, S. H\"ofner and B. Aringer}
{105}

\partline{Initial Mass Function}{}
\tocline{The stellar initial mass function}{Pavel Kroupa}
{109}
\tocline{A Possible Origin of the Mass--Metallicity Relation of Galaxies}{Carsten Weidner,
 Joachim K{\"o}ppen and Pavel Kroupa}
{120}

\partline{Stellar Population Models}
\tocline{On TP-AGB stars and the mass of galaxies}{Gustavo Bruzual A.}
{125}
\tocline{Stellar Population SEDs at 2.3\AA}{A. Vazdekis,
N. Cardiel, 
A.J. Cenarro,
J.L. Cervantes,
J. Falc\'on-Barroso,
J. Gorgas,
J. Jim\'enez-Vicente,
J.M. Mart\'{\i}n-Hern\'andez,
R.F. Peletier,
P. S\'anchez-Bl\'azquez,
S. O. Selam and 
E. Toloba}
{133}
\tocline{High resolution spectral models for solar scaled and $\alpha$-enhanced 
compositions}{P.~Coelho,
G.~Bruzual,
S.~Charlot,
A.~Weiss and 
B.~Barbuy}
{138}
\tocline{New response functions for absorption-line indices from high-resolution spectra}{R. Tantalo, C. Chiosi and L. Piovan}
{143}
\tocline{High-redshift galaxies and the TP-AGB phase}{Claudia Maraston}
{147}
\tocline{Modelling the Near-IR Spectra of Red Supergiant-dominated Populations}{Ariane Lan\c{c}on, 
 Jay S. Gallagher, 
 Richard de Grijs, 
 Peter Hauschildt,
 Djazia Ladjal, 
 Mustapha Mouhcine, 
 Linda J. Smith, 
 Peter R. Wood and
 Natascha F\"orster Schreiber}
{152}
\tocline{Stellar model choice and the MOPED fossil record}{Benjamin Panter}
{156}
\tocline{The Chemistry of the Local Group}{Brad K. Gibson}
{161}
\tocline{An optimized H$_\beta$ index for disentangling stellar clusters and galaxy ages}{Jose Luis Cervantes 
 and Alexandre Vazdekis}
{165}
\tocline{A new approach to derive [$\alpha$/Fe] for integrated stellar populations}{J.L. Cervantes, P. Coelho, B. Barbuy, and
  A. Vazdekis}
{167}
\tocline{A probabilistic formulation of evolutionary synthesis models: implications for SED fittings}{M. Cervi\~no and
 V. Luridiana}
{169}
\tocline{Synthesis models in the VO framework}{M. Cervi\~no,
E. Terlevich,  
R. Terlevich,
C. Rodrigo-Blanco,
V. Luridiana,
A. L\'opez  and E. Solano}
{171}
\tocline{Stellar Population Challenge: analysis of M67 with the VO}{M. Cervi\~no,  
 R. Guiti\'errez  and E. Solano}
{173}
\tocline{NBursts: Simultaneous Extraction of Internal
Kinematics and Parametrized SFH from Integrated Light Spectra}{Igor Chilingarian,
Philippe Prugniel, Olga Sil'chenko and Mina Koleva}
{175}
\tocline{Access to Stellar Population Models in the Virtual Observatory}{Igor V. Chilingarian}
{177}
\tocline{Simple Stellar Populations: constraints from open clusters and
binary evolution}{L. Deng and Y. Xin}
{179}
\tocline{A Binary Model for the UV-upturn of Elliptical Galaxies}{Z. Han, Ph. Podsiadlowski, A.E. Lynas-Gray and K. Schawinski}
{181}
\tocline{Comparison of different spectral population models}{Mina Koleva,
 Philippe Prugniel, Pierre Ocvirk, Damien Le~Borgne, Igor Chilingarian and Caroline Soubiran }
{183}
\tocline{Age and metallicity of Galactic clusters from full spectrum fitting}{Mina Koleva,
 Philippe Prugniel, Pierre Ocvirk and Damien Le~Borgne}
{185}
\tocline{The Effect Of Alpha-Element Enhancement On Surface Brightness Fluctuation  
Magnitudes And Broad-Band Colors}{Hyun-chul Lee, Guy Worthey, and John P. Blakeslee}
{187}
\tocline{Potential colors for studying stellar populations}{Zhongmu Li,
 Zhanwen Han and Fenghui Zhang}
{189}
\tocline{Surface-brightness fluctuations in stellar populations}{ A. Mar\' \i n-Franch
 and A. Aparicio}
{191}
\tocline{A graphical user interface for STECKMAP}{P. Ocvirk}
{193}
\tocline{Towards a calibration of SSP models from the optical to the mid-infrared}{P. Pessev, P. Goudfrooij, T. Puzia and R. Chandar}
{195}
\tocline{H$\delta$ in the integrated light of galaxies: What are we actually measuring?}
{L.C. Prochaska, J. A. Rose and R.P. Schiavon}
{197}
\tocline{Tracing stellar populations of galaxies with the SBF method }{G. Raimondo, M. Cantiello,
 E. Brocato, J. P. Blakeslee and M. Capaccioli}
{199}
\tocline{Separating Physical Components from Galaxy Spectra by Subspace Methods}{Ching-Wa~Yip, Alex~S.~Szalay, 
  Andrew~J.~Connolly and Tamas~Budav\'ari}
{201}
\tocline{Blue Stragglers in Galactic Open Clusters and Simple Stellar Population Models}{Y. Xin, L. Deng and Z.W. Han}
{203}
\tocline{Binary Stellar Population Synthesis Model}{F. Zhang,
 Z. Han and L. Li}
{205}

\partline{Stellar Populations in the Milky Way}
\tocline{Decomposition of the Galactic Disk}{Bacham E. Reddy}
{209}
\tocline{Studying Milky Way structure using stellar populations}{J.T.A. de Jong, D.J. Butler, H-W. Rix,
  A.E. Dolphin and D. Mart\'inez-Delgado}
{213}
\tocline{The HST/ACS Survey of Galactic \\ Globular Clusters}{Ata Sarajedini}
{218}
\tocline{Metallicity distribution of $\omega$ Cen Red Giants based on the Str\"omgren $m_1$
metallicity index}{A. Calamida, G. Bono, L.M. Freyhammer,
F. Grundahl, C.E. Corsi, P. B. Stetson, R. Buonanno,
M. Hilker and T. Richtler}
{223}
\tocline{Detailed Properties of Populous Clusters in the Large Magellanic Cloud
}{A.J. Grocholski, A. Sarajedini, A.A. Cole, D. 
Geisler, \\ K.A.G. Olsen, G.P. Tiede, V.V. 
Smith and C.L. Mancone}
{227}
\tocline{Mn, Cu, and Zn abundances in metal-rich globular clusters}{A. Alves-Brito, B. Barbuy and D.M. Allen}
{231}
\tocline{ New Lessons from the First Galactic Stars}{J. Andersen and B. Nordstr\"om}
{233}
\tocline{The Determination of Stellar Parameters of Giants in the Galactic
Disks and Bulge}{Joakim Bystr\"om, Nils Ryde, Sofia Feltzing, Johan
Holmberg and Thomas Bensby}
{235}
\tocline{Reconstructing the spatial distribution of the Galactic stellar halo}{M. Cignoni,
V. Ripepi, M. Marconi, J. M. Alcal\'a,
M. Capaccioli, M. Pannella and R. Silvotti}
{237}
\tocline{Membership, binarity, reddening and metallicity of red giant candidates in 
three southern open clusters}{J.J. Clari\'a, J.-C.Mermilliod , A.E. Piatti and M.C. Parisi}
{239}
\tocline{ Near-Infrared photometry of the Galactic Globular Cluster NGC 6441}{M. Dall'Ora, J. Storm, 
G. Bono, P.B. Stetson, G. Andreuzzi, R. Buonanno, F. Caputo, M. Marconi, M. Monelli, 
A. Piersimoni, V. Ripepi, L. Vanzi and A.K. Vivas}
{241}
\tocline{The NGC 2419 project:$\:$ preliminary results on stellar variability}{M. Di Criscienzo%
 C. Greco, M. Dall'Ora, V. Ripepi, G. Clementini, I. Musella, M. Marconi, L. Federici, L. Di Fabrizio, Baldacci
and M. Maio}
{243}
\tocline{Alpha-enhancement in the MW: results from the SDSS spectroscopic stellar database}{M. Franchini, 
C. Morossi,  P. Di Marcantonio and M.L. Malagnini}
{245}
\tocline{TCS-CAIN: NIR survey of the Galactic plane}{C. Gonz\'{a}lez Fern\'{a}ndez, 
A. Cabrera Lavers, F. Garz\'{o}n, P. L. Hammersley, M. L\'{o}pez-Corredoira and  B. Vicente}
{248}
\tocline{Kinematic structure in the Galactic halo at the North Galactic Pole: RR Lyrae 
and BHB stars show different kinematics}{T.D. Kinman, 
 C. Cacciari, A. Bragaglia, A. Buzzoni and 
 A. Spagna}
{250}
\tocline{The Frequency of Carbon-Enhanced Metal-Poor Stars Based on SDSS Spectroscopy}
{B. Marsteller,
 T.C. Beers, T. Sivarani, S. Rossi,  J. Knapp, B. Plez and J. Johnson}
{252}
\tocline{Structure of the Milky Way  and  the distribution of young stellar clusters.}{Maria Messineo, Karl M. Menten,
Harm J. Habing,and Monika Petr-Gotzens and Fr\'ed\'eric
Schuller}
{254}
\tocline{Integrated spectroscopy and individual spectra of stars of open cluster remnants and candidates}{D.B. Pavani,
E. Bica, A.V. Ahumada and J.J. Clari\'a}
{256}
\tocline{The Chemical Evolution of Omega Centauri}{Donatella Romano}
{258}
\tocline{A First Study of Giant Stars in the Galactic Bulge based on
    Crires spectra}{N. Ryde,
 B. Edvardsson, B. Gustafsson, and  H.-U. K\"aufl}
{260}
\tocline{The puzzling origin and evolution of stellar populations in $\omega$ Centauri}{A. Sollima,
F. R. Ferraro, M. Bellazzini and E. Pancino}
{262}

\partline{Resolved Stellar Populations in the Local Group }
\tocline{Resolved Stellar Population Modeling}{Antonio Aparicio, Sebasti\'an L. Hidalgo, Carme Gallart and Santi Cassisi}
{267}
\tocline{Star Formation History and Chemical Evolution of Resolved 
Galaxies: a New Model}{Myung Gyoon Lee and In-Soo Yuk}
{274}
\tocline{Abundances \& Abundance Ratios in our Galaxy \& the Local Group}{Eline Tolstoy}
{279}
\tocline{The Recent Star Formation Histories of Nearby Galaxies}{Evan D.\ Skillman,
John M.\ Cannon and
Andrew E.\ Dolphin}
{286}
\tocline{The ACS LCID project: overview and first results}{Carme Gallart, for the LCID Team}
{290}
\tocline{The ACS LCID Project: Quantifying the Delayed  Star Formation in Leo~A}{Andrew A. Cole
 for the LCID Team}
{295}
\tocline{The VMC survey and the SFH of some Local Group Galaxies}{Maria-Rosa L. Cioni}
{300}
\tocline{Old main-sequence turnoff photometry in the SMC: Star Formation History and Chemical Enrichment Law}{Noelia E. D. No\"el, Carme Gallart,  Antonio Aparicio,
Sebasti\'an L. Hidalgo, Ricardo Carrera, 
 Edgardo Costa and Ren\'e A. M\'endez}
{305}
\tocline{A New Deep HST/ACS CMD of I~Zw~18: Evidence for Red 
Giant Branch Stars}{A.~Aloisi, F.~Annibali, J.~Mack, M.~Tosi, 
R.~van der Marel, G.~Clementini, R.~A.~Contreras, 
G.~Fiorentino, M.~Marconi, I.~Musella and A.~Saha}
{310}
\tocline{The Star Formation History of M33's Outer Regions}{M.K. Barker, A. Sarajedini, 
D. Geisler,  P. Harding and R. Schommer}
{315}
\tocline{The ACS LCID Project:  Variable Stars in Tucana and LGS3}
{Edouard J. Bernard for the LCID Team}
{317}
\tocline{A SAGE View of the Mass Losing Sources in the Large Magellanic Cloud}{Robert D. Blum, S. Points, 
S. Srinivasan, K. Volk, M. Meixner, F. Markwick--Kemper, R. Indebetouw, B. Whitney, M. Meade, B. Babler, 
E.B. Churchwell, K. Gordon, C. Engelbracht,  B.--Q. For,  K. Misselt, U. Vijh, C. Leitherer, W. Reach,  
J.L. Hora, and The SAGE Team}
{319}
\tocline{Probing Stellar Populations in the Outskirts of NGC4244}{S. Buehler,
 A.M.N. Ferguson, M.J. Irwin, N. Arimoto and P. Jablonka}
{321}
\tocline{The Magellanic Clouds Chemical Enrichment History
via Ca II Triplet Spectroscopy}{R. Carrera C. Gallart  A. Aparicio E. Costa E. Hardy R. M\'endez and N. No\"el}
{323}
\tocline{A method for recovering the star formation history of resolved stellar populations}{M. Cignoni
 S. Degl'Innocenti, P. G. Prada Moroni and S. N. Shore}
{325}
\tocline{Integrated spectral properties of blue concentrated star clusters of the Large Magellanic Cloud}{J.J. Clari\'a, 
M.C. Parisi, A.V. Ahumada, J.F.C. Santos Jr., E. Bica and A.E. Piatti}
{327}
\tocline{A wide field survey of Sagittarius dSph. Data and tools for the study
of the Sgr Tidal Stream}{M. Correnti,
M. Bellazzini, F.R. Ferraro and L.
  Monaco}
{329}
\tocline{RR Lyrae stars in the Bootes structure}{M. Dall'Ora,
 G.Clementini, K. Kinemuchi, V. Ripepi,  M.
 Marconi, M. Di Criscienzo, L. Di Fabrizio, C.
 Greco,  C.~T. Rodgers, C. Kuehn
 and H.~A. Smith}
{331}
\tocline{Stellar populations in the Magellanic Clouds: looking through the dust}{Guido De Marchi, Nino Panagia and Martino Romaniello}
{333}
\tocline{The Stellar Structures around Disk Galaxies}{Igor Drozdovsky,
Nikolay Tikhonov,
Antonio Aparicio, Carme Gallart,
Matteo Monelli, Sebastian Hidalgo,
Edouard J. Bernard, Olga Galazutdinova and 
the LCID team}
{335}
\tocline{VLT spectrosocpy of RR Lyrae stars in the Sagittarius northern tidal stream}{S. Duffau, M. T. Ruiz, R. Zinn and A. K. Vivas}
{337}
\tocline{The star formation history of the dwarf irregular galaxy SagDIG}{
E. V. Held, 
Y. Momany, 
L. Rizzi, 
I. Saviane, 
L.~R. Bedin and
M. Gullieuszik}
{339}
\tocline{Very metal poor Classical Cepheids: variables in IZw18.}{G. Fiorentino , M. Marconi, G. Clementini
, I, Musella, A. Aloisi, F. Annibali,\\
 R. A. Contreras and M. Tosi }
{341}
\tocline{The Oosterhoff types of the Fornax dSph Globular Clusters}{C. Greco ,
G. Clementini, M.Catelan, E.Poretti, E.V.Held, M. Gullieuszik, M.Maio,
 A. Rest, N. De Lee, H.A. 
Smith and B.J. Pritzl}
{343}
\tocline{The Star Formation History of Phoenix Dwarf Galaxy using IAC-pop Algorithm}{{Sebastian L. Hidalgo, 
Antonio Aparicio 
and David Mart\'{\i}nez-Delgado}}
{345}
\tocline{The Binary Fraction of the Young Star Cluster NGC 1818
    in the Large Magellanic Cloud}{Yi Hu, Qiang Liu, Licai Deng, and Richard de Grijs}
{347}
\tocline{The outer disk stellar populations in M31}{R.\,A.\,Johnson, D.\,Faria, A.\,M.\,N.\,Ferguson and J.\,C.\,Richardson}
{349}
\tocline{A spectroscopy-based Age-Metallicity Relation of the SMC}{Andrea Kayser, Eva K. Grebel, Daniel R. Harbeck, 
Andrew A. Cole, Andreas Koch, Katharina Glatt,
John S. Gallagher and Gary S. Da Costa}
{351}
\tocline{Analysis of HST CMDs of 15 intermediate-age LMC clusters:
self-consistent physical parameters and 3D distribution}{L.O. Kerber,
B.X. Santiago and E. Brocato}
{353}
\tocline{Star Formation History of Dwarfs 
in Nearby Galaxy Groups}{L.Makarova,
D.Makarov, I.Karachentsev, A.Dolphin,
B.Tully, S.Sakai, E.Shaya, L.Rizzi, M.Sharina and
V.Karachentseva}
{355}
\tocline{A panoramic view of the Southern quadrant of the Andromeda galaxy outer halo}
{Nicolas F. Martin, Rodrigo A. Ibata  and Mike J. Irwin}
{357}
\tocline{Searching for RR Lyrae stars in the Canis Major overdensity}{C. Mateu, K. Vivas, R. Zinn and L. Miller}
{359}
\tocline{The structural complexity of the dwarf galaxies of the Local Group}
{Alan W. McConnachie, Nobuo Arimoto and Mike J. Irwin}
{361}
\tocline{{\it Spitzer} Survey of the Large Magellanic Cloud: Surveying the Agents of a Galaxy's Evolution (SAGE)}{M. Meixner, 
 K. Gordon, R. Indebetouw, B. Whitney,  M. Meade,  B. Babler, 
 J. Hora, U. Vijh, S. Srinivasan, C. Leitherer, M. Sewilo,  
 C. Engelbracht,  M. Block, B. For, R. Blum, W. Reach and J-P. Bernard}
{363}
\tocline{The Cepheids Variable Stars Population in the Local Group Dwarf Irregular Galaxy Pegasus}
{I. Meschin, C. Gallart, S. Cassisi, A. Aparicio and A. Rosenberg}
{365}
\tocline{Kinematical properties of stellar populations in the Carina dSph galaxy}
{M. Monelli,
G. Bono, M. Nonino, P. Francois , F. Th{\'e}venin, 
A. Aparicio, R. Buonanno, F. Caputo, C.E. Corsi, M. Dall'Ora, 
C. Gallart, A. Munteanu, L. Pulone, V. Ripepi, H.A. Smith, 
P.B. Stetson, A.R. Walker
and J.Q. Public}
{367}
\tocline{The ACS LCID project: data reduction strategy}{M. Monelli 
for the LCID team}
{369}
\tocline{A photometric and spectroscopic study of the stellar populations in the 
Large Magellanic Cloud}{A. Mucciarelli, F. R. Ferraro, E, Carretta,
 L. Origlia and F. Fusi Pecci}
{371}
\tocline{The Star Formation History in a SMC field: IAC-star/IAC-pop at work}{Noelia E. D. No\"el,  Antonio Aparicio, Carme Gallart, 
Sebasti\'an L. Hidalgo,
 Edgardo Costa and Ren\'e A. M\'endez}
{373}
\tocline{Stellar Populations of Halo Substructure Along The Major Axis of M31}
{J. C. Richardson, A. M. N. Ferguson, R. A. Johnson and D. C. Faria}
{375}

\partline{Stellar Populations in Early-type Galaxies }
\tocline{Integrated Spectra of Early-Type Galaxies}{James A. Rose}
{379}
\tocline{Stellar Population gradients in early-type galaxies}{Patricia S\'anchez-Bl\'azquez,
 Duncan Forbes,Jay Strader, Pierre Ocvirk,
 Jean Brodie and
 Robert Proctor}
{391}
\tocline{ Early Type Galaxies in the Mid Infrared:
a new flavor to their stellar populations}
{A. Bressan,
 P. Panuzzo,
 O. Vega,
 L. Buson, 
 M. Clemens, 
 G.L. Granato, 
 R. Rampazzo,
 L. Silva and
 J.R. Valdes}
{395}
\tocline{Stellar Populations of Decoupled Cores in E/S0 Galaxies with SAURON and OASIS}
{Richard M.\ McDermid, Eric Emsellem, Kristen L.\ Shapiro,
  Roland Bacon, Martin Bureau, Michele Cappellari, Roger L.\ Davies,
  Tim\ de Zeeuw, Jes{\'u}s Falc{\'o}n-Barroso, Davor Krajnovi{\'c}, Harald Kuntschner, Reynier F.\ Peletier and Marc Sarzi}
{399}
\tocline{The use of [Mg/Fe] to trace truncated star formation in elliptical galaxies}{Ignacio G. de la Rosa,
 Reinaldo R. de Carvalho, Alexandre Vazdekis  and Beatriz Barbuy}
{404}
\tocline{The many faces of early-type dwarf galaxies}{T. Lisker, E.~K. Grebel, B. Binggeli, M. Vodi\v{c}ka, K. Glatt,
and P. Westera}
{409}
\tocline{The star formation history of dwarf galaxies: 
First results of the MAGPOP-ITP}{Dolf Michielsen,
 Alessandro Boselli,
 Javier Gorgas,
 Reynier Peletier and the MAGPOP-ITP team}
{414}
\tocline{Central Stellar Populations of S0 Galaxies in the Fornax Cluster}{A.G. Bedregal, A. Arag\'on-Salamanca, M.R. Merrifield and N.
Cardiel}
{416}
\tocline{Galaxies with nested bars: constraining their formation scenarios}{Adriana de Lorenzo-C\'{a}ceres, Alexandre Vazdekis and J.~Alfonso L.~Aguerri}
{418}
\tocline{Stellar Populations Across cD Galaxies}{Susan I. Loubser,
A.E. Sansom and I.K. Soechting}
{422}
\tocline{Stellar population analysis of two ellipticals}{Andr\'e de C. Milone,,
Miriani Pastoriza and Mauro Rickes}
{424}
\tocline{Stellar line-strength indices distribution inside the bar region}{I.
 P\'erez, P. S\'anchez-Bl\'azquez and A. Zurita}
{426}
\tocline{IC 4200: an early-type galaxy formed via a major merger.}
{Paolo Serra, S.C. Trager, J.M. van der Hulst, T.A. Oosterloo, R. Morganti and J.H. van Gorkom}
{428}
\tocline{Stellar populations of dwarf elliptical galaxies from optical and near-IR high-resolution spectroscopic data}
{E. Toloba, J. Gorgas, A.J. Cenarro
and the MAGPOP-ITP team}
{430}

\partline{Extragalactic Globular Cluster Systems }
\tocline{Formation History of Stars and Star Clusters in Nearby Galaxies}{S. S. Larsen, M. D. Mora, J. P. Brodie and T. Richtler}
{435}
\tocline{The Globular Cluster System 
of NGC 5128}{Doug Geisler,,
Matias G\'omez, W.E. Harris, K. Woodley,
G.L. Harris, T. Puzia 
and M. Hempel  }
{440}
\tocline{Extragalactic globular clusters: unraveling galaxy formation and constraining stellar evolution theories}
{A. Javier Cenarro, Michael A. Beasley, Jay
Strader, Jean P. Brodie and Duncan A. Forbes}
{445}
\tocline{Resolving Stellar Populations in Extragalactic Globular Cluster Systems}{Maren Hempel}
{449}
\tocline{Star Cluster Population of the Interacting Galaxy System M51}{Narae Hwang and Myung Gyoon Lee}
{451}
\tocline{Planetary Nebulae in Extragalactic Young Star Clusters}{S. S. Larsen and T. Richtler}
{453}
\tocline{Ages and metallicities of Globular Clusters in M33}{Alessia Moretti and
 E. V. Held }
{455}
\tocline{Globular Cluster Systems in Massive Low Surface Brightness Galaxies}{Daniela Villegas, Markus Kissler-Patig, Andr\'es Jord\'an,  
Paul Goudfrooij and Martin Zwaan}
{457}

\partline{Stellar Populations in Late-type Galaxies }
\tocline{The Star Formation History of Late Type Galaxies}{Roberto Cid Fernandes}
{461}
\tocline{Stellar Populations in KDCs of Sa Galaxies}{Jes\'us Falc\'on-Barroso, 
Roland Bacon,
Michele Cappellari,
Roger Davies, 
P.~Tim de Zeeuw,
Eric Emsellem,
Davor Krajnovi\'c,
Harald Kuntschner,
Richard M.\ McDermid,
Reynier F.\ Peletier,
Marc Sarzi and
Glenn van de Ven}
{470}
\tocline{Nuclear Star Clusters (Nuclei) in Spirals and Connection to 
Supermassive Black Holes}{Roeland P.~van der Marel, Joern Rossa, Carl Jakob Walcher,
Torsten B\"oker, Luis C.~Ho, Hans-Walter Rix and Joseph C.~Shields}
{475}
\tocline{Stellar Populations in Spiral Galaxies}{Lauren A. MacArthur, Jes{\' u}s J. Gonz{\' a}lez and 
 St{\'e}phane Courteau}
{480}
\tocline{The Nature of Galactic Bulges from SAURON Absorption Line Strength Maps}{Reynier F. Peletier,
Jes\'us Falc\'on-Barroso, Katia Ganda, Roland Bacon, Michele Cappellari, Roger L. Davies,
P. Tim de Zeeuw, Eric Emsellem, Davor Krajnovi\'c, 
Harald Kuntschner, Richard M. McDermid, 
Marc Sarzi, and Glenn van de Ven}
{485}
\tocline{The Stellar Populations of Seyfert 2 Nuclei}{Marc Sarzi, Joseph C. Shields, Richard W. Pogge, and Paul Martini}
{489}
\tocline{Techniques for quantifying the Star Formation Morphology of Galaxies at increasing redshift.}{J. Ruyman Azzollini and J.E. Beckman}
{493}
\tocline{The edges of the stellar populations of early type spirals as probed by 
their radial brightness profiles.}{J. Beckman, L. Guti\'errez, R. Aladro, 
P. Erwin and M. Pohlen}
{495}
\tocline{A SINFONI view of the nuclear star formation ring in NGC\,613}{Torsten B\"oker, J. Falcon-Barroso, J. H. Knapen, 
E. Schinnerer, E. Allard and S. Ryder}
{497}
\tocline{Stellar Populations in the Center of the Barred Galaxy NGC~4900}{Simon Cantin, Mercedes Moll\'{a}, Carmelle Robert and Anne
Pellerin}
{499}
\tocline{Where in the Virgo Cluster are Galaxies Stripped? Stellar Population Evolution of Stripped Spiral Galaxies in Virgo}
{Hugh H. Crowl and Jeffrey D.P. Kenney}
{501}
\tocline{GHOSTS: The Resolved Stellar Outskirts of Massive Disk Galaxies}{Roelof S.\ de Jong, A.C.\ Seth, E.F.\ Bell, T.M.\
  Brown, J.S.\ Bullock, S.\ Courteau, J.J.\ Dalcanton,
H.C.\ Ferguson, P.\ Goudfrooij, S.\ Holfeltz, C.\ Purcell,
D.\ Radburn-Smith and D.\ Zucker}
{503}
\tocline{Structure and evolution of star-forming gas in late-type spiral galaxies}{Kambiz Fathi, 
John E. Beckman, 
Almudena Zurita, 
M\'onica Rela\~no, 
Johan H. Knapen, 
G\"oran \"Ostlin, 
Claude Carignan, 
Laurent Chemin, 
Olivier Daigle and
Olivier Hernandez}
{505}
\tocline{Multi-Band Bar/Bulge/Disk Image Decomposition of a Thousand Galaxies}{Dimitri Gadotti and Guinevere Kauffmann}
{507}
\tocline{News from bulges hosted by low surface brightness galaxies}{Gaspar Galaz, Alvaro Villalobos, Lorenzo Morelli, Ivan
  Lacerna, Carlos Donzelli and Leopoldo Infante}
{509}
\tocline{Two-dimensional spectroscopy of late-type spirals}{Katia Ganda, 
 Reynier F.Peletier, 
 Jes\'us Falc\'on-Barroso and Richard M. McDermid }
{511}
\tocline{Comparison of $UBVR$ photometry of giant HII regions in NGC 628 
with a detailed grid of evolution models of star clusters}{Alexander S. Gusev, Valery I. Myakutin,
Firouz K. Sakhibov  and Mikhail A. Smirnov}
{513}
\tocline{The star formation history in circumnuclear regions of galaxies}{Johan H. Knapen, Emma L. Allard, Marc Sarzi, Reynier
  F. Peletier and Lisa M. Mazzuca}
{515}
\tocline{Physical conditions of ionized gas and stellar populations in
circumnuclear starbursts}{Johan H. Knapen, Lisa M. Mazzuca and Marc Sarzi}
{517}
\tocline{Stellar population in bulge of spiral galaxies}{L. Morelli, E. Pompei, A. Pizzella, L. Coccato, E.M.Corsini,
   J. Mendez, R. Saglia, M. Sarzi and F. Bertola}
{519}
\tocline{Radial distributions of spectral absorption indices in spiral disks}{Mercedes Moll\'{a}}
{521}
\tocline{Detection of a stellar halo in NGC 4244}{A. Seth, R. de Jong, J. Dalcanton, and the GHOSTS team}
{523}
\tocline{Stellar ages and star-forming properties 
         of galaxies in a dense group around IC 65}{J. Vennik and U. Hopp}
{525}
\tocline{Sampling effects in the emission line spectra of HII regions}{M. Villaverde, V. Luridiana and M. Cervi\~{n}o}
{527}
\tocline{Stellar populations and AGN in the bulges of SDSS galaxies}{Vivienne Wild, Guinevere
  Kauffmann and Tim Heckman}
{529}
\tocline{Chemodynamical models of barred galaxies}{Herv\'e Wozniak
and L\'eo Michel-Dansac }
{531}
\tocline{Radial Dependency of Stellar Population Properties in Disk Galaxies from SDSS Photometry}{Ching-Wa~Yip and Rosemary F.~G.~Wyse}
{533}

\partline{Stellar Populations at Higher Redshifts }
\tocline{Stellar Populations at Higher Redshifts}{Tadayuki Kodama}
{537}
\tocline{Environment and the epochs of galaxy formation in the SDSS era}{D.~Thomas, C.~Maraston, K.~Schawinski, M.~Sarzi, S.-J.~Joo, 
S.~Kaviraj and S.~K.~Yi}
{546}
\tocline{Chemical clocks for early-type galaxies in clusters}{Conrado Carretero, Alexandre Vazdekis and John E. Beckman}
{551}
\tocline{A census of the physical parameters of
nearby galaxies}{Anna Gallazzi, 
 J. Brinchmann, S. Charlot and S.D.M White}
{556}
\tocline{IFU observations of the core of Abell~2218}{N. Cardiel,
S.F. S\'{a}nchez,
M.A.W. Verheijen,
S. Pedraz,
and G. Covone}
{561}
\tocline{Colors of intermediate $z$ bulges in Groth and GOODS-N}{Lilian Dom\'\i nguez-Palmero 
and  Marc Balcells}
{563}
\tocline{Stellar Population in Extremely Red Galaxies}{A.Hempel, 
   D.Schaerer, 
 J.Richard, 
  E.Egami and
  R.Pell\'o }
{565}
\tocline{Star formation properties of UV selected galaxies in the ELAIS field}{J. Iglesias-P\'{a}ramo,
V. Buat, J. Hern\'{a}ndez-Fern\'{a}ndez,
C.K. Xu, D. Burgarella and GALEX \& SWIRE teams}
{567}
\tocline{Velocity Fields of Spiral Galaxies in z$\sim$0.5 Clusters}{Elif Kutdemir, 
Bodo Ziegler and Reynier F. Peletier}
{569}
\tocline{News from z$\lesssim$6-10 galaxy candidates found behind gravitational lensing clusters}{ D.Schaerer, 
A.Hempel, 
R.Pell\'o, 
E.Egami and
J.Richard}
{571}
\tocline{Toward more precise photometric redshift estimation}{O. Vince and I. Csabai}
{573}

\partline{New Observing Facilities }
\tocline{Stellar populations -- the next ten years}{J. Bland-Hawthorn}
{577}


\tocline{Author Index}{}{591}

\newpage
\thispagestyle{plain}
\addcontentsline{toc}{section}{Preface}
\markboth{ }{ }

.
\vskip 25truemm
\section*{\Large{Preface}}
\vskip 5truemm

The last IAU Symposium on Stellar Populations was held in 1995, i.e., 
more than
ten years ago. We felt that at the time where stellar populations are 
providing
among the strongest constraints on our current picture of how galaxies 
form and
evolve an IAU Symposium to review the status of this field and to 
discuss routes
that  need to be  followed to pursue this research was largely needed. We
therefore proposed the idea to celebrate an IAU Symposium on the island 
of La
Palma, where one of the observatories that have contributed so much to the
development of this field is located, and where a new 8-10m class 
telescope will
soon see the first light. An IAU Symposium had not been held in La Palma 
since
the Observatorio de El Roque de Los Muchachos was built up about two decades
ago.\\

The Symposium was sponsored by IAU Commission 28 and supported by six IAU
Commissions relevant for the field (see acknowledgements below).  The 
Symposium
obtained the support of IAU Divisions VII and VIII and was approved by 
the IAU
Executive Committee in April 2005. The Symposium was then properly anounced
reaching a strong oversubscription with more than 300 pre-registered
participants. As a result, we were obliged to turn down a rather high
proportion of the pre-registered participants  due to the limited 
capacity of the
selected Hotel.
As a result about 40 people who were originally included in the 
waiting list were
admitted in the end. Finally, approximately 180 participants
from 25 countries attended the Symposium. IAU grants were awarded to 40
participants. Preference was given to young researchers from 
economically
less-privileged countries.

We would like to express our inmense gratitude to all the members of the LOC
for their very active involvement in the organisation of the meeting and 
above
all to Tanja Karthaus who did most of the work. We would like to 
thank the
Hotel for their great help and flexibility in all sort of aspects related to
the organisation of the event and for the participants requests. We also 
would
like to thank Sonia Soria for her great help and in particular for obtaining
most of photos during this event.  We also express our gratitude to all the
members of the SOC for their invaluable help to set up the whole scientific
programme. \\

The conference was attended by both modelers (in stellar evolution and 
stellar
populations) and observers working on Stellar Populations in the Milky Way,
the Local Group and
unresolved nearby and high redshift galaxies. We are particularly glad that
researchers of resolved stellar populations and those working with 
integrated
light, two important areas of research that have grown separately, 
attended the
Symposium to unite their efforts in the era where new ground and space
astronomical facilities are providing the means to study the same 
objects with
these two approaches. The programme included 12 reviews to introduce 
each topic
and 50 contributed talks. The SOC asked each reviewer to present a very
balanced overview of the theme. The contributed talks were selected by 
the SOC,
and special care was taken to represent all the themes
covered in the Symposium. We are particularly satisfied that a balance 
between
active senior and young researchers was reached in the programme of 
talks. In
addition, there were about 140 poster papers. Within the SOC we considered
different ways for emphasizing the true relevance of these posters. We
therefore decided to ask five experienced researchers to help by presenting
summaries of the posters, in this way giving people an overview of the
information present, so that it would be easier for the participants to 
select
the posters that they wanted to look at in detail.
Each of the five poster summary sessions was allotted 20 minutes (1 hour 
and 40
minutes in total) and all were scheduled during the first half of the 
week so
that the participants had enough time to look at the posters afterwards. The
SOC members would like to express our gratitude to S. Heap, J. Andersen, P.
Stetson, J.E. Beckman and U. Fritze, for their enthusiasm and great 
effort for
accomplishing such a difficult task.\\

We also organized two challenges, where researchers from the community
participating in the conference were asked to model a number of pre-defined
problems. The solutions were discussed, and by comparing them the audience
could get a good idea of the differences between (I) various stellar 
evolution
models and (II) different stellar population models. The specifications 
of the
proposed models and fits to be calculated had been defined in the last 
week of
June 2006 during the workshop "Fine-Tuning Stellar Population Models" 
held at
the Lorentz Center in Leiden, The Netherlands, which was attended by 
about 60
participants, 40 of which also attended the IAU Symposium 241. We would 
like to
express our personal thanks to the organizers of these challenges, A. 
Weiss and
S. Trager, for the tremendous work involved in the definition of the
challenges, for their patience and struggle with the comunity to get their
results in due time and for their efforts in summarizing the results and
conclusions. We also would like to thank all the groups who contributed to
these challenges. Finally, the Symposium included two sets of three parallel
discussion sessions, where in smaller groups people a number of
important problems in the field were discussed: 

\begin{itemize}
\item Abundance patterns of different elements in stars and galaxies: 
what are they and what can we learn from them.
\item When did the first galaxies form?
\item What can be done to make the Virtual Observatory useful for 
stellar population analysis?
\item Later stellar evolution phases and their impact on population 
synthesis
\item What is the relationship between compact ellipticals (such as 
M32)and more massive ellipticals
\item What should we do with the future instruments on the Hubble Space 
Telescope (WFC3 and COS)?
\end{itemize}

We are happy with the experience of a large conference in La Palma. The 
weather
started rough, but became very pleasant at the end of the conference. Many
participants visited the observatory, during one of the two excursions, 
and saw
the beautiful Island of La Palma. We hope that for many people this has 
been a
remarkable visit, and that many IAU Symposia will follow here. \\
\bigskip

%

\vskip\baselineskip

\noindent{\it Alexandre Vazdekis and Reynier Peletier, co-chairs SOC,\\
La Laguna, Groningen, May 5, 2007}

\newpage

\large
\mh{\centering THE ORGANIZING COMMITTEE}

\section*{Scientific}
\vskip.5\baselineskip
\noindent\begin{tabular}{@{}p{18pc}l}
A. Aparicio (Spain) & N. Arimoto (Japan)\\
B. Barbuy (Brazil) & A. Bressan (Italy)\\
G. A. Bruzual (Venezuela) & R. L. Davies (UK)\\
J. Gorgas (Spain) & T. M. Heckman (USA)\\
G. Kauffmann (Germany) & R. F. Peletier (co-chair,Netherlands)\\
J. A. Rose (USA) & D. A. Vandenberg (Canada)\\
A. Vazdekis (co-chair, Spain) & \\
\end{tabular}

\section*{Local}
\vskip.5\baselineskip
\noindent\begin {tabular}{@{}p{18pc}l}
M. Balcells & M. Beasley \\ 
J. E Beckman & E. Bejarano \\
N. Caon & C. Carretero  \\ 
J. L. Cervantes-Rodr\'{\i}guez & J. de Araoz\\
A. de Lorenzo-C\'aceres & C. Gallart \\
A. Herrero & T. Karthaus\\
N. No\"el & J. A. P\'erez Prieto\\
A. Vazdekis (chair)& \\
\end {tabular}

\section*{Acknowledgements}
\vskip.5\baselineskip
\noindent The symposium is sponsored and supported by the IAU Divisions VII
(Galactic System) and VIII (Galaxies); and by the IAU Commissions No. 28 (Galaxies)
No. 33 (Structure and Dynamics of the Galactic System),
No. 35 (Stellar Constitution),
No. 36 (Theory of Stellar Atmospheres),
No. 37 (Star Clusters and Associations),
No. 45 (Stellar Classification),
and No. 47 (Cosmology)

\vskip1.5\baselineskip

{\centering
The Local Organizing Committee operated under the auspices of the\par
Instituto de Astrof\'{\i}sica de Canarias.\par\vskip.5\baselineskip
Funding and support by the\par
International Astronomical Union,\par
Ministerio de Educaci\'on y Ciencia,\par
Universidad de La Laguna,\par
Isaac Newton Group of Telescopes (ING),\par
Patronato de Turismo de La Palma,\par
Ayuntamiento de Los Llanos de Aridane,\par
Ayuntamiento de El Paso,\par
and\par
Ayuntamiento de Bre\~na Baja.\par
are gratefully acknowledged.\par}

\end{document}